\documentclass{PoS}
\usepackage{amsmath}
\usepackage{amsfonts}

\title{
  \begin{picture}(0,0)(0,0)%
    \put(300,75){\makebox(0,0)[l]{\textnormal 
      {\normalsize KEK-CP-339, OH-HET-887}
    }}
  \end{picture}Stochastic calculation of the QCD Dirac operator spectrum with
  Mobius domain-wall fermion
}

\ShortTitle{Stochastic calculation of QCD Dirac spectrum }

\author{G. Cossu\\
  High Energy Accelerator Research Organization (KEK), 
  Tsukuba 305-0801, Japan
}
\author{H. Fukaya\\
  Department of Physics, Osaka University, Toyonaka 560-0043, Japan
}
\author{
  \speaker{S.~Hashimoto}\\
  High Energy Accelerator Research Organization (KEK), 
  Tsukuba 305-0801, Japan and\\
  School of High Energy Accelerator Science, 
  SOKENDAI (The Graduate University for Advanced Studies),
  Tsukuba 305-0801, Japan\\
  E-mail: \email{shoji.hashimoto@kek.jp}
}
\author{T. Kaneko\\
  High Energy Accelerator Research Organization (KEK), 
  Tsukuba 305-0801, Japan and\\
  SOKENDAI (The Graduate University for Advanced Studies),
  Tsukuba 305-0801, Japan
}
\author{J. Noaki\\
  High Energy Accelerator Research Organization (KEK), 
  Tsukuba 305-0801, Japan
}

\abstract{
  We calculate the spectral function of the QCD Dirac operator using
  the four-dimensional effective operator constructed from the Mobius
  domain-wall implementation. 
  We utilize the eigenvalue filtering technique combined with the
  stochastic estimate of the mode number. The spectrum in the entire
  eigenvalue range is obtained with a single set of measurements. 
  Results on 2+1-flavor ensembles with Mobius domain-wall sea quarks
  at lattice spacing $\sim$ 0.08~fm are shown.
}

\FullConference{The 33rd International Symposium on Lattice Field Theory\\
		14 -18 July 2015\\
		Kobe International Conference Center, Kobe, Japan*}

\begin{document}

\section{Introduction}
The eigenvalue density (or spectral function) of the Dirac operator
\begin{equation}
  \rho(\lambda)=\frac{1}{V}
  \langle\sum_i\delta(\lambda-\lambda_i)\rangle
  \label{eq:spectral_function}
\end{equation}
provides a probe of the spontaneous chiral symmetry breaking in QCD
through the Banks-Casher relation
$\rho(0)=\Sigma/\pi$
\cite{Banks:1979yr},
where $\Sigma$ denotes the chiral condensate
$\Sigma=-\langle\bar{q}q\rangle$
in the thermodynamical limit.
The functional form of $\rho(\lambda)$ at small $\lambda$ is
computed by one-loop
chiral perturbation theory in the $p$-regime \cite{Smilga:1993in} 
and in the mixed regime \cite{Damgaard:2008zs}, 
but the value of $\Sigma$ is to be determined by non-perturbative QCD calculations.

The most direct way of obtaining the eigenvalue density in lattice
QCD is to calculate the individual low-lying eigenvalues and to count
the number of them falling in a region suffciently close to zero.
This method was adopted in our previous works to extract the chiral
condensate in 2+1-flavor QCD
\cite{Fukaya:2009fh,Fukaya:2010na} 
using the overlap-Dirac operator.
For larger volumes, however, it becomes computationally more demanding
because of the cost and memory requirement of the Lanczos-type algorithms.

An alternative way is to stochasitically estimate the number of eigenvalues below some threshold.
It was first implemented in \cite{Giusti:2008vb} for this particular
problem. 
In this work we introduce a variant of this method to calculate the
spectral function.
Namely, we utilize the Chebyshev filtering technique combined with a
stochastic estimate of the mode number.
As described in the next section, the method is more flexible and can
be used to calculate the whole spectrum at once.
We use the lattice ensembles generated with 2+1 flavors of the Mobius
domain-wall fermion at a lattice spacing $a\simeq$ 0.08~fm.

\section{Chebyshev filetering}
One can evaluate the number of eigenvalues in an interval $[a,b]$ of a
hermitian matrix $A$, which is supposed to be $D^\dagger D$ of any
lattice Dirac operator $D$, as
\begin{equation}
  n[a,b] = \frac{1}{N_v}
  \sum_{k=1}^{N_v}\xi_k^\dagger h(A) \xi_k
  \label{eq:n}
\end{equation}
with Gaussian random vectors $\xi_k$, 
which has a normalization
$(1/N_v)\sum_{k=1}^{N_v}\xi_k^\dagger\xi_k=12V$ in the limit of 
large $N_v$, the number of random vectors.
$h(A)$ is a function of matrix $A$ that works as a filter of
eigenvalues. 
Without $h(A)$, (\ref{eq:n}) simply counts the total mode number.
By preparing $h(x)$ returning 1 in the range
$[a,b]$ and 0 elsewhere, we may stochastically count the number of
modes in that interval. 
The statistical error is given by a square-root of the mode number in
$[a,b]$.
When the number of eigenvalues in the range $[a,b]$ is
sufficiently large, $N_v=1$ could already give a precise estimate.

One can use the Chebyshev polynomial $T_j(x)$ 
to approximate the filter $h(x)$:
\begin{equation}
  h(x) = \sum_{j=0}^p g_j^p \gamma_j T_j(x).
\end{equation}
The coefficients $\gamma_j$ and $g_j^p$ are known numbers fixed
once the interval $[a,b]$ is given.
The conventional Chebyshev minmax approximation is obtained with
$\gamma_j$, while the Jackson stabilization factor $g_j^p$ is
introduced to suppress the oscillation typical in the Chebyshev
expansion \cite{DiNapoli:2013}.
In order for the Chebyshev approximation to work, the whole
eigenvalues of $A$ have to be in the range $[-1,1]$.

After the ensemble average (over gauge configurations) one obtains
\begin{equation}
  \bar{n}[a,b] = \frac{1}{N_v} \sum_{k=1}^{N_v}
  \left[
    \sum_{j=0}^p
    g_j^p \gamma_j
    \langle\xi_k^\dagger T_j(A) \xi_k\rangle
  \right].
\end{equation}
An important observation is that 
once the stochastic estimates of $\langle\xi_k^\dagger T_j(A)\xi_k\rangle$
are calculated for each $j$ the eigenvalue count in any interval
$[a,b]$ can be obtained by combining them with the corresponding
coefficients $g_j^p\gamma_j$.
Namely, the interval can be adjusted afterwards, independent of the
costly calculation of the polynomial of the matrix $A$.
Details of the method are found in \cite{DiNapoli:2013}.

The Chebyshev polynomial can be easily constructed using the
recurrence formula
\begin{equation}
  T_0(x)=1, 
  \;\;\;
  T_1(x)=x,
  \;\;\;
  T_j(x) = 2x T_{j-1}(x)-T_{j-2}(x).
\end{equation}
One can also use the relations
$T_{2n-1}(x)=2T_{n-1}(x)T_n(x)-T_1(x)$ and
$T_{2n}(x)=2T_n^2(x)-T_0(x)$,
in order to reduce the numerical efforts.
With infinitely large $p$ the filtering function $h(x)$ is exactly
reproduced; 
at finite $p$, the approximating function is smeared around the
borders $a$ and $b$, inducing a systematic error.

\begin{figure}[tbp]
  \centering
  \includegraphics[width=9.5cm,clip=true]{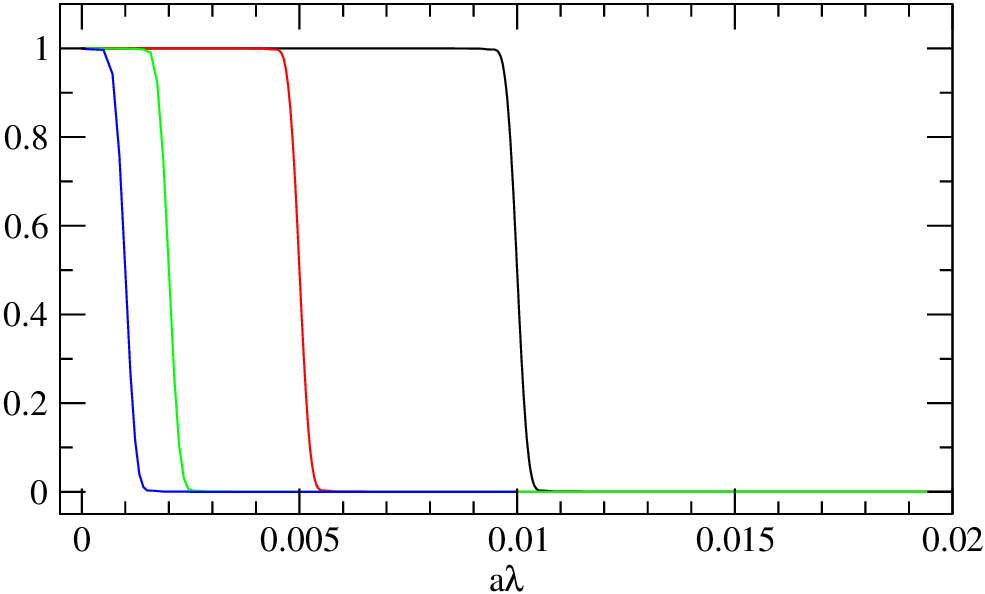}
  \caption{
    Step function approximation given by the Chebyshev
    polynomial at order $p=8000$.
    Typical results for the interval $[0,\delta]$ = 0.01, 0.005, 0.002
    and 0.001 are shown from right to left.
  }
  \label{fig:step}
\end{figure}

For the four-dimensional effective Dirac operator of domain-wall
fermion, the eigenvalues of $D^\dagger D$ are in the range
$[0,1]$.
In Figure~\ref{fig:step}, we plot the step functions for the interval $[0,\delta]$ of the eigenvalue of $|D|$ with
$\delta$ = 0.01, 0.005, 0.002 and 0.001.
(For the correspondence between the eigenvalues of $D^\dagger D$ and
those of $|D|$, see below.)
The polynomial order is fixed to $p$ = 8000.
One can see that the step function is well approximated away from the
boundary.
Near the boundary, the edge is rounded off. Its effect is relatively
more important for smaller $\delta$.
The error estimated for the area, which has to be $\delta$, 
is 
0.8\% for $\delta=0.01$ and
1.5\% for $\delta=0.005$, scaling as $1/\delta$.

\section{Lattice calculation}
The JLQCD collaboration has generated a new set of ensembles of
2+1-flavor QCD with Mobius domain-wall fermion for sea quarks.
It aims at achieving good chiral symmetry, {\it i.e.} the residual
mass is order 1~MeV or smaller.
Three lattice spacings are chosen as
$1/a$ = 2.45, 3.61 and 4.50~GeV, 
which allow well controlled continuum extrapolation even including
charm quarks as valence quarks.
The up and down quark masses correspond to the pion mass $M_\pi$ of
230, 300, 400 and 500~MeV; 
two strange quark masses are chosen such that they sandwich the
physical value. 
Lattice volume is $32^3\times 64$, $48^3\times 96$ and 
$64^3\times 128$, depending on the lattice spacing,
and the physical volume satisfies the nominal condition $M_\pi L\gtrsim 4$.
This set of ensembles have been used for a variety of
applications
\cite{Noaki:2014sda,Fukaya:2015ara,Tomii:2015exs,Nakayama:2015hrn,Fahy:2015xka}.

In this preliminary work, we use the coarse lattice ($1/a$ = 2.45~GeV)
of size $32^3\times 64$, out of the above mentioned ensembles.
The number of configurations is 50 for each ensemble, taken out of
10,000 HMC trajectories.
The number $N_v$ of the Gaussian noise vector $\xi_k$ is 1.
The up and down quark masses in the lattice unit are $am_{ud}$ =
0.019, 0.012, 0.007 and 0.0035.

We calculate the eigenvalue density of the hermitian operator 
$D^{(4)\dagger}D^{(4)}$ made of the four-dimensional (4D) effective
operator
\begin{equation}
  D^{(4)}=[P^{-1}(D^{(5)}(m=1))^{-1}D^{(5)}(m=0)P]_{11}.
\end{equation}
Here, $D^{(5)}(m)$ represents the five-dimensional (5D) Mobius
domain-wall operator with mass $m$.
For the eigenvalue count we took $m=0$, {\it i.e.} the massless Dirac
operator. 
The 4D effective operator is constructed by multiplying the inverse of
the Pauli-Villas operator ($m=1$) and taking the 4D surfaces
(represented by the subscript ``11'')
appropriately projected onto left- and right-handed modes by a
projection operator $P$.
See, for instance, \cite{Boyle:2015vda} for more details.

For each application of $D^{(4)}$ on 4D vectors, we have to calculate
the inverse of the Pauli-Villars operator, for which the conjugate
gradient iteration of order 40--50 is involved.
Although the inversion is much less expensive than the calculation of
light quark propagator, the total numerical cost is substantial
because we have to multiply $D^{(4)\dagger}D^{(4)}$ $p$-times.
($p$ = 8000 in this analysis.)

The eigenvalues of $D^{(4)\dagger}D^{(4)}$ are in the region $[0,1]$,
and we rescale the operator as $A=2D^{(4)\dagger}D^{(4)}-1$ to match
the region of the Chebyshev approximation.
Since the effective 4D operator satisfies the Gisparg-Wilson relation
very precisely, we assume that eigenvalues of $D^{(4)}$ lie on a
circle in the complex plane.
In the following, the eigenvalue $\lambda$ stands for that projected
onto the imaginary axis as
$\lambda=\sqrt{\lambda_{D^{(4)\dagger}D^{(4)}}/
(1-\lambda_{D^{(4)\dagger}D^{(4)}})}$.

\begin{figure}[tbp]
  \centering
  \includegraphics[width=10cm,clip=true]{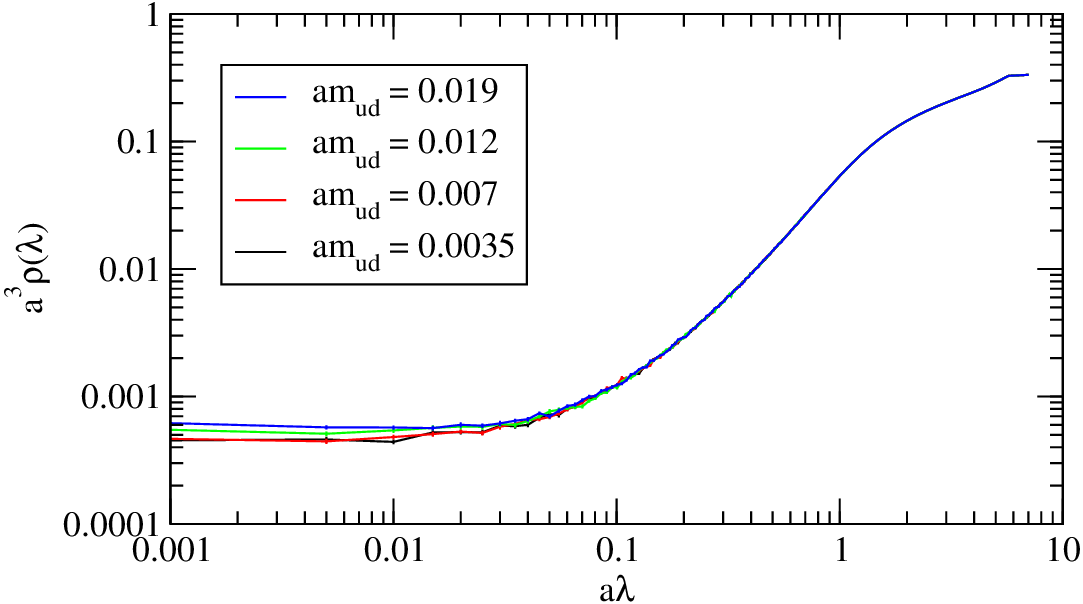}
  \caption{
    Spectral function in a logarithmic scale.
    The lattice data are plotted for four values of up and down quark
    masses.
  }
  \label{fig:spect_log}
\end{figure}

Figure~\ref{fig:spect_log} shows the eigenvalue spectrum for the whole
range of $\lambda$ in the lattice unit.
Both axes are in a logarithmic scale.
For each bin of $[a,b]$, it is constructed as
$\rho(\lambda;\delta)=(1/2V)\bar{n}[a,b]/\delta$
with a bin size $\delta$.
(Therefore, it satisfies 
$\lambda=\sqrt{a/(1-a)}$ and $\lambda+\delta=\sqrt{b/(1-b)}$.)

One can clearly see that the number of eigenvalues increases toward
higher $\lambda$ and saturate at some point of $O(1)$ due to the
discretization effect, which should otherwise behave like
$\sim\lambda^3$ for asymptotically large $\lambda$.
There is no visible quark mass dependence in this region.
On the lowest end, it approaches a constant corresponding to
$\rho(0)$, from which one extracts the chiral condensate.

The same data are plotted in Figure~\ref{fig:spect} in a linear scale.
The individual bin has a width of $\delta$ = 0.005.
With this binsize, the systematic error due to the Chebyshev
approximation is well below the statistical error.

\section{Analysis using $\chi$PT formula}
\begin{figure}[tbp]
  \centering
  \includegraphics[width=10cm,clip=true]{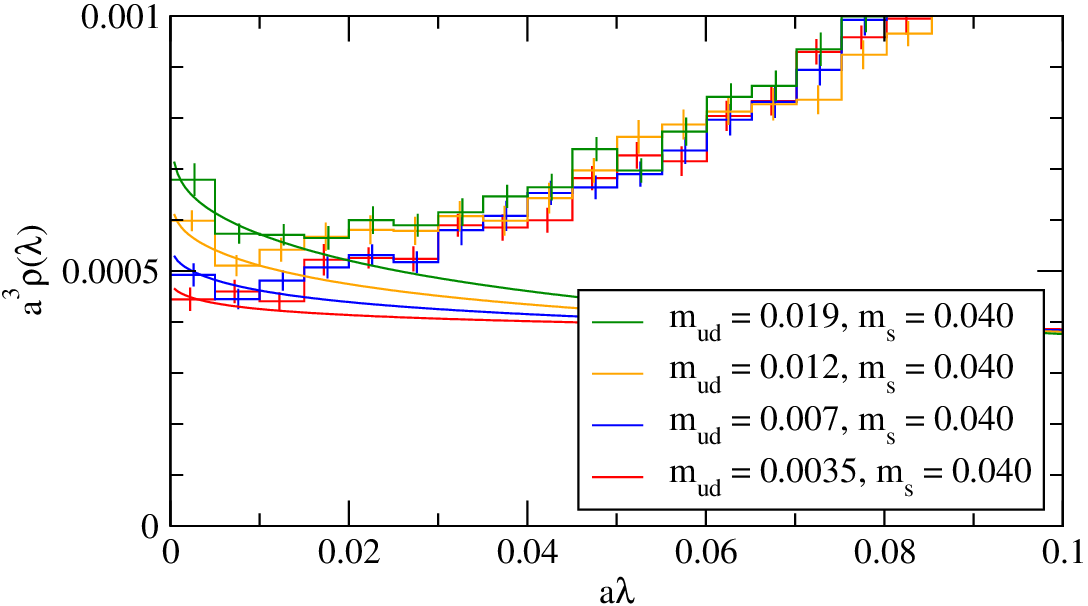}
  \caption{
    Spectral function in a linear scale.
    The binsize is taken as $\delta$ = 0.005.
    The lattice data are plotted for four values of up and down quark
    masses (histogram with four different colors).
    Curves are those of one-loop chiral perturbation theory.
  }
  \label{fig:spect}
\end{figure}

Figure~\ref{fig:spect} shows a clear dependence of $\rho(0)$ on the
sea up and down quark mass $m_{ud}$.
Namely, $\rho(0)$ gets lower for smaller $m_{ud}$.
Furthermore, a peak develops near $\lambda=0$ for heavier sea quarks.
Qualitatively, it is understood as the effect that the fermion
determinant is no longer active below $\lambda\lesssim m_{ud}$ to
suppress the near-zero modes.
More near-zero eigenvalues may then survive for larger $m_{ud}$.

In order to obtain the chiral condensate $\Sigma$, one has to take the 
thermodynamical limit, {\it i.e.} the infinite volume limit and
then the massless quark limit.
The order of the limits is crucial; $\rho(0)$ vanishes in the massless
limit on any finite volumes.
Fortunately, such volume and mass dependences are well understood in
chiral perturbation theory ($\chi$PT), and we may identify the volume
beyond which the system is effectively in the large volume limit.
All our lattices satisfy that condition, and we use the $p$-regime
$\chi$PT formula in the following analysis.

One-loop formula for $N_f=2$ is available in \cite{Smilga:1993in}.
It is written in terms of the leading order low-energy constants (LEC) 
$\Sigma$ and $F$ as well as the next-to-leading order LEC $L_6$.
The pion decay constant controls the size of the next-to-leading order
corrections.
In this preliminary analysis, we fix it to a nominal value $F$ =
90~MeV.

We fit the value of
$\bar{\rho}[0,\delta]=(1/\delta)\int_0^\delta d\lambda\rho(\lambda)$
with the one-loop $\chi$PT expression with $\delta=0.01$.
Namely, both the lattice data and one-loop $\chi$PT are integrated in
the same region.
This value of $\delta$ corresponds to the scale of pion mass of
$\delta\Sigma/F^2\simeq$ 300~MeV, for which one expects that
one-loop $\chi$PT converges reasonably well.

\begin{figure}[tbp]
  \centering
  \includegraphics[width=10cm,clip=true]{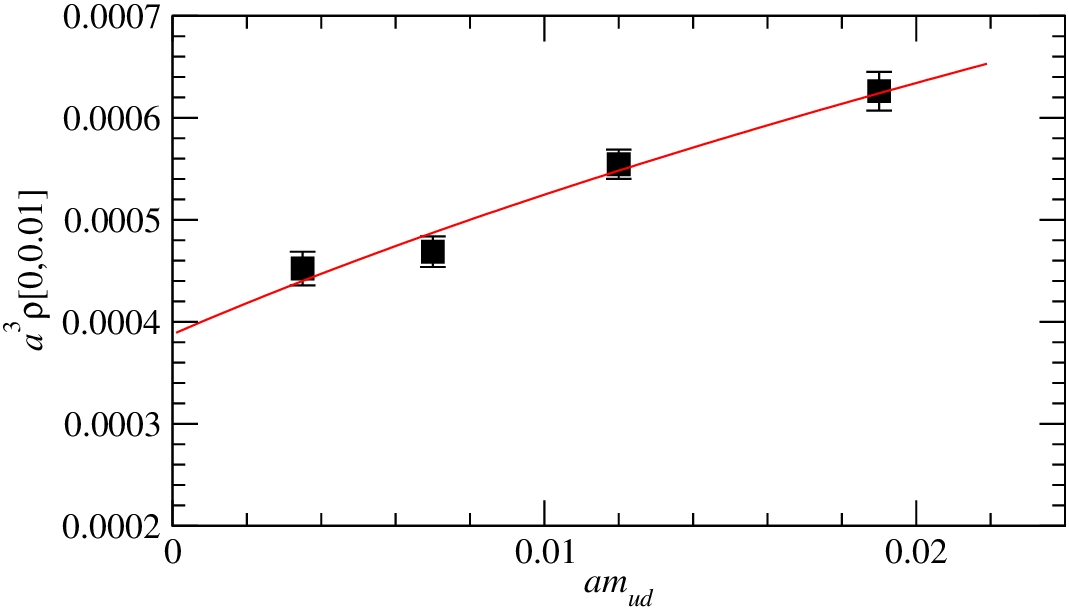}
  \caption{
    Chiral extrapolation of the partially integrated spectral function
    $\rho[0,\delta]$ with $\delta=0.01$.
    The one-loop $\chi$PT curve at $N_f=2$ is shown together with the
    lattice data (black square).
  }
  \label{fig:chi}
\end{figure}

Chiral extrapolation of $\rho[0,0.01]$ is shown in
Figure~\ref{fig:chi}.
The one-loop $\chi$PT curve shows a slight curvature due to the chiral
logarithm.
The fit yields 
$\Sigma^{1/3}$ = 262.0(1.7)~MeV and
$L_6$ = 0.00031(7) with $\chi^2/\mbox{dof}$ = 1.13.
With these parameter values, we draw the curves of $\rho(\lambda)$ for
each quark mass in Figure~\ref{fig:spect}.
They explain the rise near $\lambda=0$ for larger quark masses, but
the data beyond $a\lambda\sim$ 0.015 cannot be explained by one-loop
$\chi$PT. 

Renormalizing to the $\overline{\mbox{MS}}$ scheme at 2~GeV, we obtain
$[\Sigma(\mbox{2~GeV})]^{1/3}$ = 260.0(1.7)~MeV, where we use the
renormalization factor $Z_S(\mbox{2~GeV})$ determined from the
analysis of short-distance current correlator \cite{Tomii:2015exs}.

Probably because the strange quark is too heavy to apply one-loop
$\chi$PT, a fit with the 2+1-flavor $\chi$PT formula failed to
reproduce the lattice data.

\section{Discussions}
The Chebyshev filtering technique allows precise evaluation of the
eigenvalue count in a sufficiently small bin to calculate the
eigenvalue spectrum.
The method is especially suitable for the 4D effective operator of the domain-wall fermion since the eigenvalue of $D^{(4)\dagger}D^{(4)}$ is
limited in $[0,1]$.
For the Wilson fermion, the range is $[0,64]$ (in the free theory),
and one needs much higher order polynomial to obtain the same
precision.
This would nearly compensate the numerical effort to construct the
(expensive) 4D effective Dirac operator from domain-wall fermion.

This preliminary analysis has been done using partial data out of the
full data set at three lattice spacings and various sea quark masses.
We plan to include the data on finer lattices that allow us to
extrapolate to the continuum limit.

\vspace*{5mm}
We are grateful to Julius Kuti for private communications on the
technique employed in this work.
Their own work was also presented at this conference.
Numerical calculation was performed on the Blue Gene/Q supercomputer
at High Energy Accelerator Research Organization (KEK) under a support
of its Large Scale Simulation Program (No. 14/15-10).
The code set Iroiro++ \cite{Cossu:2013ola}, which is highly optimized
for Blue Gene/Q, is used.
This work is supported in part by JSPS KAKENHI Grant Number 25800147,
26247043 and 15K05065.

\end{document}